# Toward Understanding the Conditions that Promote Higher Attention in Software Developments – A First Step on Music and Standups

Rozaliya Amirova
Innopolis University
Innopolis, Russia
r.amirova@innopolis.ru

Sergey Masyagin
Innopolis University
Innopolis, Russia
s.masyagin@innopolis.ru

Anastasia Reprintseva Innopolis University
Innopolis, Russia
a.repryntseva@innopolis.ru

Giancarlo Succi
Innopolis University
Innopolis, Russia
g.succi@innopolis.ru

Herman Tarasau
Innopolis University
Innopolis, Russia
h.tarasau@innopolis.ru

## ABSTRACT

Nowadays, Computer Science tightly entered all spheres of human activity. To improve quality and speed of development process, it is important to help programmers improve their working conditions. This paper proposes a vision on exploring this issue and presents in conjunction a factor that has been claimed multiple time to affect the effectiveness of software production, concentration and attention of software developers. We choose to focus on developers brain activity and features that can be extracted from it.

## 1 INTRODUCTION

This paper presents a vision on starting studying more systemati cally aspects related to software development that are important and yet remain often unexplored. Questions like "In which envi ronment should we code?" or "Should we listen to music while we debug?" or also "Should we stand up during meetings?" and so on are rarely the focus of research investigations. Still, they have anecdotal potentials to affect the outcome of our work significantly. This paper proposes a vision on exploring this issue and presents in conjunction a factor that has been claimed multiple time to af fect the effectiveness of software production – concentration and attention of software developers.

We choose to focus on developers brain activity and features that can be extracted from it, and present the first steps that were done to explore the changes of attention of software developers in different situations. We are convinced that the exploration of that vision deserves greater attention and that it has a potential of further application in the studies of developers productivity and a possibility to be adopted in real software production processes. Our first step towards the vision was to explore the methods that can be used to measure developers attention and then examine two environments that can promote higher concentration: standing up on meetings and coding with music.

The paper is organized as follows: Section 2 explores the back ground behind the vision, Section 3 states the methods that can be used to measure developers attention based on their brain activity and collected EEG signals. Sections 4 and 5 present the preliminary findings of two experiments based on the methods discussed in the previous section, and finally, we provide the discussion of the vision in Section 6 and conclude and outline the work in Section 7.

## 2 BACKGROUND

Software is a product of the mental activity of developers. Their mental activity is highly influenced by external factors, such as sounds, illumination, air humidity, etc. Creating better conditions for developers could potentially increase their performance and well-being.

In this work, we promote a vision that developers' brain activity can tell us about their productivity at work and study which conditions result in higher productivity.

As a primary instrument, the Electroencephalography(EEG) was chosen. EEG is a non-invasive, portable method of brain activity monitoring. In last decades EEG is gaining popularity. Different studies used EEG to determine the attention level of subject per forming learning task[1], to understand if a person is meditating or concentrated[2] and find the alertness level in real-time[3]. This work investigates environment scenarios: programming with music and stand-up meetings.

Music is the external factor that may greatly influence the soft ware developer's state of mind. An example, the relaxing music can decrease the anxiety of particular person[4]. There is a lot of peo ple, who are listening to music while working. Software developers are not the exception. Understanding how the music influence the attention level may be the key to create better working conditions for developers.

Another area where the attention of software developers is cru cial is meetings. An essential part of the success of the organisations

Rozaliya Amirova, Sergey Masyagin, Anastasia Reprintseva, Giancarlo Succi, and Herman Tarasau

is effective communication and interaction among all levels – devel opers, management and stakeholders. But despite meetings being a necessary part of many organisations everyday life, as many as half of them are rated as "poor" and lead to a waste of more than 200 billion US dollars per year [5]. The success of a meeting is affected by the attention of its individual participants. The productivity and performance of the team may suffer if one developer is not attentive or concentrated, which can indirectly result in the decreased quality of the product and customer satisfaction.

Since in our study, we focus on software developers, later we will stick to analysing their levels of attention and emphasise their roles on the meetings and while listening to music.

Indeed, the root of this work is on metrics and empirical software engineering, for which a significant amount of research has been published in the last three decades, such as [6–34].

## 3 MEASURING ATTENTION

Preparing to study this topic, we conducted a Systematic Literature Review (SLR) [35] in which we analysed the modern methods of measuring attention. According to this work, electroencephalogram (EEG) is one of the most promising method of attention monitoring for developers, because it provides accurate data, as it reads their functional state of the brain with a high frequency (of the order of a thousand hertz) and allows researches to conduct experiments in real conditions during programming, since it is portable. There are different methods of analyzing the level of attention using EEG, for example:

- Scanning the level of Alpha and Theta frequencies [36] [37] [38] [39]
- Measuring the level of Event-related Desycronization (ERD) [40]
- Alpha wave coherence [41] [42] [43]

Regarding the first one, Theta power tends to increase during dif ficult task relative to a simple task, whereas power of alpha band increases in the simple task compared to difficult tasks [36] [44] [45] [37]. Event-Related Desynchronization (ERD) measures how much neuron populations no longer synchronously react after be ing triggered to perform the given task [40]. More difficult tasks cause bigger ERD difference between resting and working samples. The ERD is equal to the percentage of change of power band be tween the resting period before a working sample and the working sample itself. Authors of [39] report the following conclusions:

- Lower alpha band desynchronization indicates an attention • Upper alpha band desynchronization indicates semantic mem ory performance
- Theta band synchronization indicates episodic memory and the processing of new information

Coherence is a technique that can catch how two or more elec trodes relate to each other and how synchronised their activation is. It detects the interrelationships between different parts of the brain by capturing the synchronisation of electrical activity in the brain and between different electrodes. In [46], authors investigate the coherence of Alpha activity during a Sustained Attention to Response Task (SART). Sustained attention can be characterised as a physiological state that describes the readiness to detect or respond to rear signals over a long time period. In [47, 48] authors performed imaging studies which showed that the activation of frontal and parietal lobes mostly of the right hemisphere indicate sustained attention performance.

## 4 ATTENTION IN STANDUP MEETINGS

To understand which external stimuli affect the attention level of developers on a meeting, we conducted an experiment that mea sured developers' biophysical signals on meetings. To capture that we used signals from a a portable 24-channel wireless EEG Smart BCIcap, provided by Mitsar company.

The key idea behind that was to collect the data about brain activity during a standup versus sitdown meeting. We asked four developers that adopt the Agile software development lifecycle and participate in team meetings to take part in our experiment [49–51]. Each subject participated twice: once while they were standing up during the meeting and once while they were sitting.

### 4.1 Methodology

The recording and the analysis of the data were done with the help of EEG Studio software, where it can be automatically inter preted as wavebands. Since each band is associated with a set of mental activities specific for each band, we focused on the ones that are associated with attention and concentration tasks the most. Namely, on Theta frequency all over the cortex, but mainly in the frontal lobe and the suppression of Alpha band in occipital, parietal and posterior temporal brain regions, which indicates the rise of attention.

Data processing consisted of two steps: data processing, where we applied filters to the raw data and divided it into epochs and feature extraction, where we extracted specific data features for further analysis. First of all, we selected only a subset of EEG channels (electrodes) for the analysis. It was done due to several reasons:

- to optimize the performance and reduce the computational complexity – the sessions of the experiment were long, re sulting in a lot of data samples
- to exclude the noise that cannot be cleared with filters – frontal electrodes are strongly affected by face muscle move ment and eye activity (blinking, moving)

For the analysis the following channels were selected:

- frontal lobe – F7, F3, Fz, F4, F8
- occipital lobe – O1, O2
- parietal lobe – P3, Pz, P4
- posterior temporal lobe – T6, T5

Electrode Cz was also included in the analysis since it was chosen as a reference electrode for EEG recording.

An amplitude filter was applied to remove the artefacts, and a low-pass/high-pass filter was used to filter out all waves, except Alpha, Theta and Beta.

The most important part of the experiment is to extract the correct features that can tell us about the levels of attention during meetings. For this experiment, we chose the following features:

(1) mean Alpha, Theta and Beta signal of the selected channels (2) Alpha wave coherence

Toward Understanding the Conditions that Promote Higher Attention in Software Developments – A First Step on Music and Standups

Table 1: Differences of the mean values among all partici pants

| Band | Stand up | Sit down | Sit down - Stand up |
|---|---|---|---|
| Theta | 6,6 | 6,56 | 0,04 |
| Alpha | 5,96 | 8,16 | -2,2 |
| Beta 1 | 2,2 | 2,73 | 0,53 |
| Beta 3 | 3,11 | 3,14 | 0,03 |
| Theta / Alpha | 1,43 | 1,41 | -0,02 |

## 4.2 Results

We automatically detected and removed all the artefacts with EEG Stidio software, since they can introduce random variance in further analysis. A part of the recording was excluded if an amplitude of one the channels was higher than a 200♦♦V. In order to keep only the bands we are interested in and extract each individual band and subband from the data, we applied one-pass, zero-phase, non-causal bandpass (low-pass and high-pass) filters to the data.

To compare the levels of attention during stand up versus sit down meetings, we divided the data into two respective epochs and analysed how much of each waveband was present in each channel. To do so, we applied spectral analysis of the power of each band and built tables for each epoch. Then we grouped them by the participant. Those tables were used to calculate the mean values of each band and then compared the values for stand up meetings and sit down meetings.

To undestand how the difference in the mean amount of rhythmic (power) frequency reflects attention, we need to state the following:

- More Theta waves appear with increasing task difficulty, dur ing activation of working memory and information uptake. Thus, the decrease indicates that a participant becomes more relaxed and undergoes less hard mental activity.
- The increase in Alpha band shows the state of mental rest, and the reduction points out to a stronger mental activity, focused attention, engagement and readiness to pick up the information.

Hence, taking those facts into account, we observed that 3 out of 4 participants showed a reduction in Alpha activity (2,2% on average) during stand up meetings which can indicate a more focused and attentive state of mind.

There was also a pattern of increasing Theta activity in 3 out of 4 cases, but the increase was insignificant (0,04% on average), so we cannot say for sure that this indicates a higher level of attention. The table with the differences if the mean values among all participants is shown in table 1.

Another technique that we decided to use for data analysis is coherence analysis. The metric that was used is the Coherence Coefficient (CC), ranging from 0 to 1, one being the highest value indicating very high coordination of brain regions.

A table of Coherence Coefficients and a Coherence map was built for each data sample. An example of a Coherence map for stand up meeting and for a sit-down meeting is illustrated in figure 1.

Figure 1: Coherence maps for meetings

# 5 ATTENTION WHILE PROGRAMMING WITH MUSIC

Rozaliya Amirova, Sergey Masyagin, Anastasia Reprintseva, Giancarlo Succi, and Herman Tarasau

For subjects 1 and 2, the ERD of all alpha bands for music is higher than control. But for subject 5 we may see the opposite

Music treatment was chosen as primary topic, as it may influence the work of programmers. The main idea was to compare treat ment(music) and control(no music) experiments. Seven male pro grammers participated in experiments. Each subject participated twice: for music and no-music experiment.

## 5.1 Methodology

The EEG studio was used to record the brain activity. The placement of electrode was done according to 10-20 international schema. In order to analyze the data, two techniques were used: Alpha/Beta ratio and Event-related desynchronization. As stated in [52], there exists a correlation between Alpha and Beta wavebands. For ex ample, Alpha activity shows that the brain is in a relaxed state, whereas Beta activity is mostly related to stimulation [53].

Data processing can be described in two steps:

- Pre-processing: applying filters to raw data. This step is per formed via EEG Studio Processing software. Amplitude fil tering for [-200��V ; 200��V] was applied. Then Notch filtering was used to remove the AC noise from signal.
- Feature extraction: Alpha/Beta ratio and Event-related desyn chronization were used to get features: First, The range of filters was used to get particular bands: L1A, L2A, UA, Theta and Beta. After that, Short-Time Fourier transform was used to get time distribution of each sub-band.

Then, mean values for each channel and setup was calculated.

## 5.2 Results

The resulting ERD and Alpha/Beta ratio values are given below: Table 2: ERD and Alpha/Beta ratio for music

|  | L1A(music) | L2A(music) | UA(music) | Theta(music) | AB_Ratio(music) |
|---|---|---|---|---|---|
| subject1 | 0.298 | 0.402 | -0.014 | 1.542 | -1.044 |
| subject2 | 0.535 | 0.409 | -0.181 | -1.38 | 1.918 |
| subject3 | -0.051 | 0.326 | -0.572 | -0.47 | -0.192 |

(a) sit down meeting

(b) stand up meeting

The goal that we wanted to reach by analyzing the coherence of waves is to detect a pattern of Alpha activity in frontal and parietal lobes of the right hemisphere.

Most of the experiments showed cohesion of Alpha band in parietal, occipital and temporal lobe and the coherence of Alpha in parietal and frontal lobes had higher CC in the frontal lobe during stand up meetings. Also, the coherence of parietal to temporal lobe was lower during stand up meetings.

From these results, we observed that there indeed were patterns of Alpha band activation in frontal and parietal lobes for some participants and some showed a higher activation during stand-ups versus sit down meetings, which can indicate that they maintained a higher sustained attention during the meeting and were able to better respond to the contents of the meeting.

Further investigation of this technique should be done to omit the possibility of the observed being caused by regular standing activity. One should make sure that alpha activation in the frontal lobe relates to attention specifically.

|  |  |  |  |  |  |
|---|---|---|---|---|---|
| subject4 | 0.988 | 0.511 | 1.865 | 0.478 | -0.09 |
| subject5 | -1.905 | -2.038 | -1.078 | 0.271 | -0.363 |
| subject6 | 0.135 | 0.39 | -0.02 | -0.441 | -0.229 |

Table 3: ERD and Alpha/Beta ratio for control

|  | L1A(control) | L2A(control) | UA(control) | Theta(control) | AB_Ratio(control) |
|---|---|---|---|---|---|
| subject1 | -0.175 | -0.366 | 0.18 | -0.294 | -1.323 |
| subject2 | 0.044 | -0.851 | 0.198 | -0.529 | 1.524 |
| subject3 | 1.431 | -1.079 | -1.373 | -0.331 | -0.125 |
| subject4 | -1.352 | 0.05 | 0.087 | 0.128 | 0.729 |
| subject5 | -0.713 | 0.648 | 1.599 | -0.899 | -0.569 |
| subject6 | 0.764 | 1.598 | -0.691 | 1.924 | -0.235 |

The evaluation of the Alpha/Beta ratio was done by comparing the results of the treatment and control experiment. As a result, the Alpha/Beta ratio is higher for music experiment, than for control. These results may show that people are becoming more attentive during listening to music. On the other hand, as there is a little amount of data, thus there are the preliminary considerations. The music could just interfere with the brainwaves and make the signal noisier.

situation. The same holds for Theta ERD band values. For some subjects (2, 4, 6) the ERD band is desynchronized, and for others not.

In conclusion, it is quite difficult to interpret the result of experiments now. We may state that music somehow influences the Alpha/Beta ratio, but the meaning of the influence is not clear.

## 6 DISCUSSION ON THE VISION

Current work promotes a vision that developers' brain activity can tell us about their productivity at work and study which conditions result in higher productivity. We conducted two experiments to investigate whether the higher level of concentration could be achieved if participants sit or stand during a Scrum meeting. The second experiment explores whether it is better to program with or without music.

First of all, we performed a Systematic Literature Review [35] on measuring the level of attention and investigated that one of the most accurate, but yet applicable in a real environment is EEG because it captures small changes in a functional state of the brain and is portable. In current work, we discuss different approaches of analysis of the EEG data. Then two experiments were conducted. First experiments, showed a reduction in Alpha activity (2,2% on average) during stand up meetings which can indicate a more focused and attentive state of mind compared to sit-down. Analysis of Alpha band cohesion showed greater sustained attention during the stand-up meeting compared to sit-down, which indicated the ability to better respond to the contents of the meeting. During the second experiment, we analysed the impact of listening to music while programming and got contradictory results. For some participants, ERD of all alpha bands with music is higher than control, which indicated the lower level of attention. But for some of them, the results were the opposite. The same holds for Theta ERD band values.

The results are still at a preliminary stage, and with our vision, we encourage the further study of the different conditions that promote greater attention of software developers. Future studies can involve the application of other methods not mentioned in this paper, extending the dataset or exploring other conditions that can affect the attention.

## 7 CONCLUSIONS

It is important to study the conditions of developers work to help them work more productively and concentrate better. In current work, we conducted experiments and got a preliminary hypothesis that stand-up meetings can help to increase the level of attention compared to sit-down. The second experiment showed that listening to music during programming for some people can help, whereas for others not. However, more experiments should be conducted to get the final results. Moreover, we plan to use techniques from machine learning and computational intelligence to gather a better understanding of the underlying phenomena [54, 55]. This study shows that the functional state of the brain can be a clue to identify optimal working conditions for programmers in order to achieve maximum productivity. Studying factors which impact the research.

level of concentration of developers can help programmers work better and timely detect problems related to burnout.

## ACKNOWLEDGEMENTS

We thank Innopolis University for generously supporting this future directions," 01 2001.

Rozaliya Amirova, Sergey Masyagin, Anastasia Reprintseva, Giancarlo Succi, and Herman Tarasau